\documentclass[referee]{mn2e}

\usepackage{graphicx}
\usepackage{amsmath}
\usepackage{url}

\title[Neutrinos from Novae]
{High energy neutrinos from fast winds in Novae}
\author[W. Bednarek \& A. \'Smia\l kowski]
{W. Bednarek \& A. \'Smia\l kowski\\ 
University of \L \'od\'z, Department of Astrophysics, Faculty of Physics and Applied Informatics,
ul. Pomorska 149/153, 90-236 \L \'od\'z, Poland,\\
wlodzimierz.bednarek@uni.lodz.pl, andrzej.smialkowski@uni.lodz.pl\\}
\begin{document}

\date{Accepted . Received ; in original form }

\pagerange{\pageref{firstpage}--\pageref{lastpage}} \pubyear{2015}

\maketitle

\label{firstpage}

\begin{abstract}
We discuss a scenario in which TeV neutrinos are  produced during explosions of Novae.
It is argued that hadrons are accelerated to very high energies in the inner part of a Nova 
wind,
as a result of reconnection of the strong magnetic field of a White Dwarf. Hadrons are expected to 
interact efficiently with a dense matter of the wind, either already during the acceleration process or 
during their advection with the equatorial wind.
We calculate the neutrino spectra, and estimate the muon neutrino event rates in the IceCube telescope, 
in the case of a few Novae. In general, those event rates are unlikely to be detected with the 
present neutrino detectors. However, for favourable location of the observer, some neutrino events might be 
detected not only from the class of Novae 
recently detected in the GeV $\gamma$-rays by the {\it Fermi}-LAT telescope
but also from novae not detected in $\gamma$-rays. The GeV $\gamma$-ray emission observed from Novae 
cannot originate in terms of the model discussed here, since
protons are accelerated within a few stellar radii of the White dwarf, i.e. in 
the region in which GeV $\gamma$-rays 
are expected to be severely absorbed in the interactions with the radiation field and the matter of the wind.  
\end{abstract}

\section{Introduction}

Recently, GeV $\gamma$-ray emission has been detected by the {\it Fermi}-LAT telescope from explosions of 
several Novae 
(Abdo et al.~2010, Ackermann et al.~2014, see also Franckowiak et al.~2018 and the recent review by 
Chomiuk et al.~2021). It has been shown that in the case of a few Novae, GeV $\gamma$-ray emission cannot 
extend with a simple 
power-law spectrum through the TeV energy range (Ahnen et al.~2015). However, the recurrent, 
symbiotic Nova RS Oph shows $\gamma$-ray emission extending up to sub-TeV energies  (Wagner et al.~2021).
Those $\gamma$-ray observations clearly indicate that the high energy processes play a very important role at 
the early phase of Nova explosion, i.e between a few to several days up to a few weeks after initial optical 
outburst. 

In fact, acceleration of particles, up to TeV energies in the Nova shocks,
has been expected already before these discoveries (see Tatischeff \& Hernanz 2007). However localization of 
the emission region and 
the type of involved radiation processes are not clear up to now. Two mechanisms for the $\gamma$-ray 
production are usually considered, i.e. interaction of hadrons with the background matter and electrons with 
the optical radiation (Abdo et al.~2010, Sitarek \& Bednarek~2012, Martin \& Dubus~2013, 
Metzger et al. 2015, Ahnen et al.~2015, Vurm \& Metzger~2018, Martin et al.~2018).

According to the present model of the Nova explosion, the thermonuclear runaway leads to ejection of 
a part of the layer of matter from the White Dwarf (WD) surface. The matter of the envelope, with the mass 
in the range $\sim 10^{-7} - 10^{-3}$~M$_\odot$ (Gallagher \& Starrfield 1978, Shore~2012), moves with 
the velocity of the order of a thousand km s$^{-1}$. However, nuclear reactions still continue in 
the rest of the layer turning to production of a fast wind from the hot surface of the WD. 
The radiation is emitted on the level close to (or even larger than) the Eddington limit (Kato \& Hachisu~1994; 
Friedjung~2011). This fast wind, with the velocity of a few thousand km s$^{-1}$, reaches slower expanding  
envelope in a few days, forming a strong shock (Li et al.2017, Aydi et al.~2020a). 

Production of $\gamma$-rays in hadronic processes, occurring in the external shocks, should be also 
companied by the emission of neutrinos with comparable energies (Razzaque et al.~2010). 
However, no significant neutrino excess could be identified from the direction of promising galactic source candidates and no significant excess was identified for the entire emission of the Galactic plane
(Aartsen et al.~2017, Abbasi et al.~2021, Albert et al.~2018).
In the case of Novae, neutrinos with GeV energies will be very difficult to detect with the present detectors, 
unless the  
$\gamma$-ray spectra do extend through the TeV energy range (Metzger et al.~2016, Fang et al.~2020). 
Neutrinos, with the TeV energies, are also expected provided that both electrons and hadrons are accelerated. 
However, hadrons reach clearly larger energies due to less efficient cooling (Sitarek \& Bednarek~2012). 
In fact, the first upper limits on the TeV neutrino fluxes from the Nova RS Oph  has been already reported by
the IceCube Collaboration (Pizzuto et al. 2021).

Here we investigate another scenario in which the high energy neutrinos are produced by hadrons accelerated 
in the 
inner, optically thick part of the fast wind as a result of reconnection of the magnetic field of strongly 
magnetized WD. 
We argue that such additional mechanism can accompany the process of acceleration of particles in 
the wind/ejected envelope shock which is likely responsible
for the observed GeV $\gamma$-ray emission. In fact, $\gamma$-ray emission occurs simultaneously 
with the optical emission which is expected to be powered by the shocks (Aydi et al. 2020b).

\begin{figure}
\vskip 7.truecm
\includegraphics{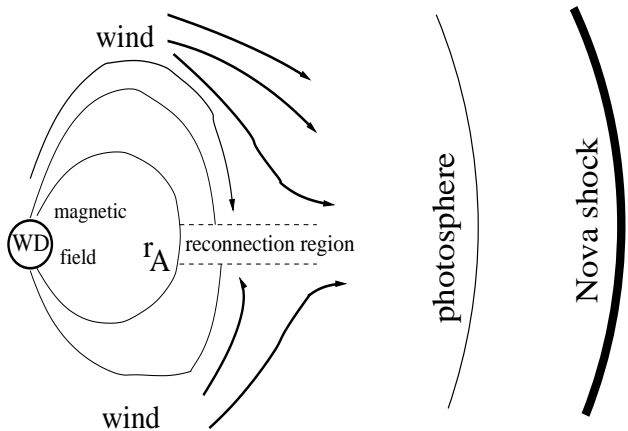}
\caption{Schematic representation of the model (not to scale) for the neutrino production in the inner 
region of the WD wind. We consider the binary systems with the  parameters for which the energy density of 
the WD magnetic field  dominates 
over the energy density of the wind close to the WD surface. Then, a part of the magnetosphere, close to 
the WD, has the dipole structure. However, at some distance, i.e. at the Alfven radius $r_{\rm A}$, the wind 
starts  to dominate, turning the magnetic field into reconnection. We argue that hadrons can be efficiently 
accelerated to the TeV energies in this reconnection region along the equatorial part of the wind. 
Relativistic hadrons are next captured in the random component of the turbulent wind. They are isotropised and 
slowly advected along the equatorial wind region. Hadrons suffer strong energy losses on the inelastic 
collisions with the matter of the wind producing pions. Neutrinos, produced from decay of charged pions, 
escape through the WD photosphere and the Nova ejecta without absorption.}
\label{fig1}
\end{figure}
\section{Model for the wind in Nova}

We consider the standard model for the Nova in which thermonuclear explosion occurs in the thick layer 
of matter on the surface 
of the WD. The layer of matter appears as a result of the accretion process from 
the companion star as observed in the Cataclysmic Variables. Since the WDs are characterised by a strong 
surface magnetic field (up to $\sim 10^8$ G), the accretion process, and the distribution of 
the matter on the WD surface, can be strongly inhomogeneous.
During the explosion, a significant amount of 
the matter (of the order of $10^{-5} - 10^{-3} {\rm M}_\odot$) is expelled with the velocity which is 
typically below $10^3$ km~s$^{-1}$ (see the review by Chomiuk et al. 2021). 
The surface of the WD is heated to temperature of the order 
of $\sim 10^6$~K. The nuclear reactions still continue in a layer of matter on the WD surface. 
Produced thermal radiation, with luminosity in the range 
$10^{38-39}$ erg~s$^{-1}$, can even exceed the Eddington luminosity for the WD. Due to the pressure of this 
radiation, a fast wind with the velocity of the order of $\sim 2000-4000$ km s$^{-1}$ is launched.  The kinetic energy of the wind is comparable to 
the thermal luminosity of this fast wind. It is expected that the fast wind can be aspherical due to 
the inhomogeneous layer on the WD surface and also strong magnetic field of the WD. 

In fact, WDs are characterised by strong surface magnetic fields of the order of $\sim 10^7$ G in the case of
isolated objects (Wickramasinghe \& Ferrario 2000). In the case of accreting WDs of the AM Her type the average 
surface magnetic fields are $(38\pm 6)\times 10^6$ G (Wickramasinghe \& Ferrario 2000). The surface magnetic 
fields of 
the intermediate polars are expected to be comparable. For example, recent observations of the  
Nova Herculis 2021 show modulation of the soft X-ray emission which is interpreted as due to the presence of
a strong magnetic field (Drake et al.~2021).
The initial structure of the magnetic field of the WD is expected
to be of the dipole type. During initial ejection of the WD envelope, the dipole structure of the magnetic field 
is modified by the dense moving plasma. However, during the phase of  a fast and less dense wind, the magnetic 
field can recover its dipole structure
starting from the WD surface. As a result, the magnetosphere of the WD 
can be characterised by a complex structure. It is similar to that postulated in the case of the early 
type massive stars with fast winds (Usov \& Melrose 1992, Trigilio et al. 2004, Leto et al. 2006, 
Leto et al. 2017, Bednarek~2021). In general, two regions can be distinguished.
Below the Alfven radius $R_{\rm A}$, where the energy density of the magnetic field dominates over 
the kinetic energy density of the wind, the magnetosphere recovers its dipole structure. 
The winds launched from the Northern and Southern region, driven by the magnetic fields
towards the equatorial plane, collide with each other forming a plasma with temperature of the order of 
several $10^6$ K (Shore \& Brown 1990, Babel \& Montmerle 1997, Gagne et al. 2005). The magnetic field 
lines in this part of  the magnetosphere are closed.   
Above $R_{\rm A}$, the energy density of the wind dominates over the energy density of the 
magnetic field. In such a case, the magnetic field is forced into reconnection in the equatorial region 
of the magnetosphere. 
Particles can be accelerated in this reconnection region to relativistic energies. 
The geometry of this scenario is shown in Fig.~1.

Here, we are interested in the radiation processes due to the protons accelerated  
in this reconnection region. The maximum energies of protons depend on the length scale 
of the reconnection region. Accelerated protons already interact with the matter 
either already in the acceleration region or after being
injected into the equatorial region of the WD wind. In the second case, they can be effectively 
isotropised, provided that the wind shows significant level of turbulence.
After injection to the equatorial wind, relativistic protons are slowly advected in the outward direction 
with the equatorial wind, suffering multiple interactions in a dense wind relatively close to the WD surface.
We calculate the fluxes of neutrinos, produced in collisions of relativistic protons with the wind matter, 
and consider whether they can be detected by the present and future
neutrino detectors. Note that, neutrinos, produced during the acceleration process in the reconnection region, 
are expected to be collimated along the equatorial wind. On the other hand,
protons, escaping from the reconnection region into the equatorial wind, produce neutrinos 
in a much larger solid angle.

\section{Acceleration of hadrons in the WD wind}

We consider the acceleration region in the fast wind of a Nova. 
The Alfven radius in the windy magnetosphere of the WD can be estimated from the balance,
$B^2/8\pi = \rho v_{\rm w}^2/2$, where $\rho$ is the density of the wind and $v_{\rm w} = 3\times 10^3v_3$ km~s$^{-1}$ 
is the WD wind velocity. 
The magnetic field strength of the WD is of the dipole type, $B = B_{\rm s}/r^3$,
where $R = r R_{\rm WD}$ is the distance from the WD and $B_{\rm s} = 10^7B_7$ G is 
the magnetic field strength on the surface of WD at the equator. 
We assume the radius of the WD  $R_{\rm WD} = 6\times 10^8$ cm  (Nauenberg~1972). 
Density of the matter is given by the WD mass loss rate,
\begin{eqnarray}
\rho = {{2L_{\rm w}}\over{4\pi 
r^2R_{\rm WD}^2v_{\rm w}^3}}\approx 10^{18}{{L_{38}}\over{v_3^3r^{2}}}~~~{\rm cm^{-3}}.
\label{eq1}
\end{eqnarray}
\noindent
The mass loss rate, $\dot{M}$, determines the power of the wind
$L_{\rm w} = \dot{M}v_{\rm w}^2/2 = 10^{38}L_{38}$ erg~s$^{-1}$. Then, the Alfven radius 
(expressed in units of the radius of the WD) is located at,
\begin{eqnarray}
r_{\rm A} = [R_{\rm WD}^2B_{\rm s}^2v_{\rm w}/(2L_{\rm w})]^{1/4}\approx 2.7 [B_7^2v_3/L_{38}]^{1/4}.
\label{eq2}
\end{eqnarray}   
The energy density of the wind dominates over the energy density of the magnetic field in the whole region 
above the surface of the WD for the condition $B_7 < 0.13(L_{38}/v_3)^{1/2}$. 
Then, the wind expands freely everywhere above the surface of the WD.
However, when the magnetic field of the WD is stronger, then the structure of the wind changes.
Above the distance estimated in Eq.~2, the magnetic field of the WD wind can be forced into the reconnection
in the equatorial region of the dipole magnetic field (see Fig.~1). 
In such a reconnection region, particles can reach relativistic energies. Since the 
wind is expected to be turbulent, the reconnection region is characterized by some degree
of coherence. We assume that the dimension of the reconnection region is proportional
to its distance from the WD, i.e. $L_{\rm rec} = \eta r R_{\rm WD}$, where the scaling factor
$\eta = 1\eta_{1}$ has to be less than unity.  

The maximum energies of protons accelerated in the reconnection region can be estimated from,
\begin{eqnarray}
E_{\rm max} = e\beta_{\rm w}\alpha B L_{\rm rec}\approx  1.8v_{3}\eta_{1}B_7 r^{-2}~~~{\rm PeV},
\label{eq3}
\end{eqnarray}
\noindent
where $\beta_{\rm w} = v_{\rm w}/c = 0.01v_3$ is the velocity of the wind in units of the velocity 
of light, $\alpha$ is the reconnection efficiency of the order of $\alpha\sim 0.1$ (e.g. Uzdensky~2007), 
and $e$ is the elemental charge. At the bottom border of the reconnection region, determined by 
the Alfven radius $r_{\rm A}$, the maximum energies of particles are 
\begin{eqnarray}
E_{\rm max}\approx  240\eta_{1}(v_3L_{38})^{1/2}~~~{\rm TeV}.
\label{eq4}
\end{eqnarray}
\noindent
Already during the acceleration process, hadrons might be able to lose energy on the pion production in 
collisions
with the dense radiation and matter of the wind. We estimate the importance of those processes on the final 
energies of hadrons.

The electromagnetic luminosity of a typical Nova is in the range, 
$L_{\rm WD} = 10^{38} - 10^{39}$ erg~s$^{-1}$. 
We assume that the Nova emission is close to the Eddington luminosity of the White Dwarf, i.e. 
$\sim 10^{38}$ erg~s$^{-1}$ is emitted from the surface of the White Dwarf. Then, the surface temperature has 
to be  $T_{\rm WD}\approx 8\times 10^5$~K (see Kahabka \& van den Heuvel~1997 and also the recent case of 
the Nova RS Oph, Pei et al. 2021). 
The energies of thermal photons are equal to $\varepsilon = 3k_{\rm B}T_{\rm WD}\sim 200$~eV. The density of
these photons is calculated assuming that the emission from the WD surface is of the black body type 
with temperature $T_{\rm WD}$. It is equal to $n_{\rm ph}\approx 10^{19}r^{-2}~~~{\rm cm^{-3}}$.
The minimum energies of protons, required for production of pions in collisions with such thermal 
photons, have to be 
$E_{\rm p}^{\rm min} = E_{p\gamma\rightarrow \pi}^{\rm th} m_{\rm p}c^2/\varepsilon\sim 700$~TeV, 
where $E_{p\gamma\rightarrow \pi}^{th} = 140$~MeV is the threshold energy for the pion production in 
collision of the $\gamma$-rays with the proton. 
We conclude that protons, accelerated in the reconnection region, are not able to lose 
energy in collisions with dense radiation field due to too low energies.

On the other hand, the mean free path for the energy losses of relativistic protons, in collisions with 
the matter of the wind, is
\begin{eqnarray}
\Lambda_{\rm pp} = (\kappa_{\rm pp}\rho \sigma_{\rm pp})^{-1}\approx 6.7\times 10^7v_3^3r^2/L_{38}~~~{\rm cm},
\label{eq5}
\end{eqnarray}
where the inelasticity coefficient is $\kappa_{\rm pp} = 0.5$. This mean free path is shorter than
the length scale of the reconnection region, $L_{\rm rec} =\eta R_{\rm WD}r = 6\times 10^8\eta_1r$~cm,
at distances $r < 10\eta_1 L_{38}/v_3^3$.
Then, the maximum energies of protons due to the energy losses are 
\begin{eqnarray}
E_{\rm max}^{\rm pp} = e\beta_{\rm w}\alpha B \Lambda_{\rm pp}\approx  200v_{3}^4B_7/(L_{38}r)~~~{\rm TeV}.
\label{eq6}
\end{eqnarray}
\noindent
For the bottom edge of the reconnection region at $r_{\rm A}$ (see Eq.~2), the maximum energies of protons, 
due to the energy losses, are
\begin{eqnarray}
E_{\rm max}^{\rm pp}\approx  74v_{3}^{15/4}B_7^{1/2}/L_{38}^{3/4}~~~{\rm TeV}.
\label{eq7}
\end{eqnarray}
\noindent
We conclude that the maximum energies, to which protons can be accelerated, are given by Eq.~4
for $\Lambda_{\rm pp} < L_{\rm rec}$ (acceleration limited by the energy losses on pion production), 
and by Eq.~7 (for the acceleration not limited by the energy losses). 
These maximum energies ($E^{\rm p}_{\rm max}$) are reported for a few 
considered example Novae in Table.~1.  The maximum energies are calculated for the surface magnetic field of 
the WD equal to $10^7$ G and $10^8$ G (values in brackets).
The scaling factor of the reconnection region is assumed to be $\eta = 1$.

\section{Production of neutrinos}

Here we show the results of a simple estimation of the muon neutrino event rates in the neutrino telescope, 
in terms of the above defined model, in the case of protons injected with the mono-energetic spectrum. 
Next, a more detailed numerical calculations, in the case of acceleration of protons with the power-law 
spectrum (spectral index -2), are presented. These two extreme models are investigated since  
acceleration of particles in the reconnection regions is expected to produce flat power-law spectra. 
The calculations of the neutrino spectra are done applying simulated
spectra of muons and pions produced by relativistic protons at a given energy using
CORSIKA Monte Carlo package with the QGSJET-II model for the high-energy interactions (Heck et al. 1998).

\subsection{Quasi-monoenergetic injection of protons}

At first, we provide {\bf very} simple estimates of the neutrino event rates in the IceCube telescope by assuming that, 
particles escape from the reconnection region into the equatorial wind. They are isotropised 
due to the turbulence in the wind. The Larmor radius of protons, with energies given by Eq.~7, in the local 
magnetic field is equal to 
$R_{\rm L}\approx 4.4\times 10^5v_3^{9/2}/L_{38}^{3/2}B_7$~cm. It is clearly shorter than the typical 
distance scale considered in the model, $rR_{\rm WD}$. Therefore, particles are frozen into the wind being 
slowly advected with the velocity of the equatorial 
wind. Due to a very large density of the wind in this region (see Eq.~1), we assume that all 
injected relativistic protons interact producing pions. 
We estimate analytically the number of muon neutrinos in a very simple way by 
assuming that two mono-energetic muon neutrinos are produced from the decay 
of a single pion. On the other hand, the number of pions, from a single mono-energetic 
proton-proton 
interaction, is given by the multiplicity $\mu$. Every muon neutrino takes on 
average one forth of the energy of produced pion and the average number of produced
pions depends on their multiplicity (which depends on the energy of interacting proton)
and on the inelasticity coefficient. So then, in the case of the interaction of 
mono-energetic protons we do not take into account the spectrum of produced secondary 
particles. However, it is taken into account in our more realistic model for protons
injected with the power-law spectrum (see next section).
The energies of neutrinos, from decay of those pions, can be roughly estimated from,
\begin{eqnarray}
E_\nu\sim \kappa_{pp}E_p/(4\mu)\approx 0.36v_{3}^{15/4}B_7^{1/2}/L_{38}^{3/4}~~~{\rm TeV}, 
\label{eq8}
\end{eqnarray}
where the inelasticity coefficient $\kappa_{\rm pp} = 0.5$, 
the average multiplicity of the charged pion production in collisions of protons with the matter changes slowly 
with the proton energy. It is equal to $\mu\approx 25.8$ for the protons with the 
characteristic energies given by Eq.~7 (see Grosse-Oetringhaus \& Reygers~2010). 
When estimating neutrino energies, we do not take into account energy losses of pions and muons
before their decay to neutrinos. In fact, They are expected to be negligible in the case of charged pions 
(due to their short decay time scale) but might be important in the case of muons. In such a case, 
the predicted fluxes of muon neutrinos should be reduced by a factor of two.  
We calculate the number of muon neutrinos from, $N_\nu\sim 2\mu N_p$,
where the total number of accelerated protons is $N_{\rm p} = L_{\rm p}/E_{\rm p}$, $L_{\rm p}$ is 
the total energy in accelerated protons and $E_{\rm p}$ is the energy of mono-energetic protons. We assume 
that protons take a part of the wind energy which fall onto the surface of 
the reconnection region in the range of distances between $r_{\rm A}$ and $2r_{\rm A}$. This surface  
defines a part of the solid angle, $\Omega = 0.1\Omega_{-1}$, of the wind ejected from the WD surface. 
If the efficiency of proton  acceleration is of the order of 
$\chi = 0.1\chi_{-1}\%$, then $L_{\rm p} = 0.01 \chi_{-1}\Omega_{-1}E_{\rm w}$. 
The typical lifetime of $\gamma$-ray emission from Novae is of the order of $\tau = 10\tau_{10}$ days. 
The total wind energy can be estimated on 
$E_{\rm w} = 10^{38}L_{38}\tau\Omega\sim 8.6\times 10^{42}L_{38}\tau_{10}\Omega_{-1}$ ergs, and 
the energy in  relativistic protons is 
\begin{eqnarray}
L_{\rm p} =  8.6\times 10^{41}L_{38}\chi_{-1}\tau_{10}\Omega_{-1}~~~{\rm ergs}.  
\label{eq8b}
\end{eqnarray}
Then, the total number of emitted neutrinos is,
\begin{eqnarray}
N_\nu\sim 2\mu N_{\rm p}\sim 3.7\times 10^{41}\chi_{-1}\Omega_{-1}\tau_{10}L_{38}^{7/4}/(B_7^{1/2}v_3^{15/4})~~~{\rm \nu},
\label{eq9}
\end{eqnarray}
\noindent
The neutrino flux on the Earth from the example Nova, which exploded at a typical distance of 3 kpc, is
\begin{eqnarray}
F_\nu = {{N_\nu}\over{4 \pi D^2}}\sim  
3.8\times 10^{-4}{{\chi_{-1}\Omega_{-1}\tau_{10}L_{38}^{7/4}}\over{B_7^{1/2}v_3^{15/4}D_3^{2}}}~~~{\rm \nu~cm^{-2}}. 
\label{eq10}
\end{eqnarray}
\noindent
The number of neutrino events produced by such a Nova in the IceCube type neutrino detector is estimated as,
\begin{eqnarray}
N_{\rm det}^\nu = SF_\nu\approx  0.38{{\chi_{-1}\Omega_{-1}\tau_{10}L_{38}^{7/4}}\over{B_7^{1/2}v_3^{15/4}D_3^{2}}}~~~{\rm \nu/Nova}, 
\label{eq11}
\end{eqnarray}
where we used the effective detection area of the IceCube with a Deep Core, 
$S\approx 10^3$~cm$^{2}$ at $E_\nu = 0.36$~TeV (Abbasi et al~2012). 
The above simple estimate of the neutrino event rate from the example Nova is done for the lower limit on 
the wind power and the energy conversion efficiency from the WD wind to relativistic
protons equal to 1$\%$. In such a case, the IceCube neutrino detector has a chance to observe a few neutrino 
events from the whole population of Novae, which are already detected in the GeV $\gamma$-rays by 
the {\it Fermi}-LAT telescope.
In case of no detection, the parameters of the considered model such as, the acceleration efficiency 
of protons $\chi$, the solid angle of the wind which falls into reconnection region, or the activity period of 
the energetic wind $\tau$, will be constrained.

In the next subsection, we also consider more realistic scenario in terms of the proposed model in which 
protons are injected from the reconnection region with the power-law spectrum. They cool completely 
in the dense WD wind. In this case we make use of a more detailed numerical calculations.

\subsection{Protons with the power-law spectrum}

In this scenario, we calculate the muon neutrino event rates assuming complete cooling of relativistic protons, 
since the column density of the matter in the WD wind is very large. This column density is estimated on 
$X_{p\gamma} = \int_{\rm r_{\rm A}}^{\infty} \rho(R) R_{\rm WD}dr\sim 350L_{38}/v_3^3$~gram. Relativistic protons 
are advected with the wind with a relatively low velocity,
$v_{\rm w}\ll c$. Therefore, they have time to cool completely in the interactions with the matter. Protons
are injected with the power-law spectrum from the reconnection region up to the maximum energies as estimated 
in Sect.~3.1. The spectrum of protons is normalized to its power $L_{\rm p}$ (see Eq.~9),
$L_{\rm p} = A \int_{E_{\rm min}}^{E_{\max}}E_{\rm p}^{-\beta}E_{\rm p} dE_{\rm p}$, where $E_{\rm max}$ is 
the maximum energies of protons (see Eq.~4 and Eq.~7) and $E_{\rm min}$ is equal to 10 GeV. In this way, 
the normalization coefficient 'A'is obtained. 

The muon neutrino fluxes expected on the Earth from a few example Novae (see their parameters reported 
in Table~1)
are shown in Fig.~2. Those neutrino spectra are compared to the Atmospheric Neutrino Background (ANB) 
within $1^\circ$ of the source (Lipari~1993).
We also calculate the muon neutrino event rate in the IceCube neutrino telescope, for two example models of 
proton acceleration, following the formula,
\begin{eqnarray}
N_\nu = \int_{E_\nu^{\rm min}}^{E_\nu^{\rm max}} S(E_\nu) {{dN_\nu}\over{dE_\nu}}dE_\nu,
\label{eq..}
\end{eqnarray}
\noindent
where $S(E_\nu)$ is the effective area of the IceCube neutrino detector as a function of neutrino energy 
(Abbasi et al.~2012) and $dN_\nu/dE_\nu$ is the neutrino spectrum produced by protons in collisions with 
the matter.

\section{Specific examples of $\gamma$-ray Novae}

We performed example calculations of the expected muon neutrino flux from a few Novae 
detected in 
the GeV $\gamma$-rays by the {\it Fermi}-LAT. These Novae differ in basic parameters 
(e.g. the total kinetic energy) allowing acceleration of protons to different maximum 
energies. Thus, they are different representative cases from the Nova population.
The parameters of these Novae, needed for 
the modelling, are collected in Table.~1. 
In our calculations, we estimate the lower 
limit on the total energy of the Nova explosion as a product of the power of the wind 
and duration of 
$\gamma$-ray emission from Nova. The optical bolometric luminosity of 
the Nova ($L_{\rm Bol}$) is estimated as the ratio of the $\gamma$-ray power and 
the coefficient describing the ratio of the $\gamma$-ray 
emission to the observed optical bolometric emission 
(the parameter $\xi$ in Table~1). 
The results of observations in the optical energy range of the novae considered by us in this work
are reported in Aydi et al. (2020b) and Abdo et al.~(2010). The maximum energies, to which 
protons can be  accelerated, are estimated in Eq.~4 (in the case of unsaturated 
acceleration) and by Eq.~7 
(in the case of saturation of the acceleration process by proton energy losses in 
collisions with the matter). 
In the case of unsaturated acceleration, 
protons are ejected from the reconnection regions into the equatorial wind. They are isotropised producing 
neutrinos in the whole solid angle. In the case of saturated acceleration, neutrinos are mainly produced in 
the general plane of the reconnection region (see Fig.~1), i.e.
in a relatively small solid angle. In this case
neutrinos are emitted anisotropicaly. For the Novae, V407 Cyg, V906 Car, and V339 Del, we show 
the expected spectra of neutrinos (Fig.~2) and their event rates in the IceCube neutrino telescope (Tab.~1). 
It is assumed that $1\%$ of the Nova total energy ($E_{\rm Nova}$) is converted to relativistic protons 
($L_{\rm p} = 0.01E_{\rm Nova}$), with the power-law spectrum and spectral index equal to $\beta = 2$.
This value for the spectral index has been selected since the modelling of the acceleration of 
reconnection regions postulate the spectra of particles with the spectral index even lower than 2 
(e.g. Guo et al. 2016). Since we do not know what is the detailed geometry of the reconnection region, 
and also the collimation of produced 
neutrinos, the event rates of neutrinos are calculated assuming isotropy. Note, that in the case of a significant 
collimation of protons, interacting already during the
acceleration process in the reconnection region, and favourite location of the observer, the neutrino event 
rates can be enhanced.   

\begin{figure}
\vskip 7.truecm
\includegraphics{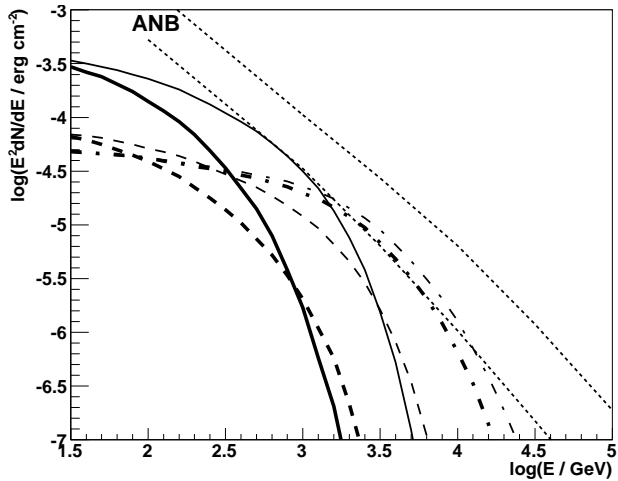}
\caption{The spectra of muon neutrinos from the wind regions of a few Novae: V906 Car (solid curves), 
V407 Cyg (dot-dashed), and V339 Del (dashed). The parameters of the Novae are listed in Table.~1.  
It is assumed that 1$\%$ of the energy of the Nova wind is transferred to relativistic protons. 
Two models, for the surface magnetic field, equal to $10^7$ G and $10^8$ G, are considered for the estimation
of the maximum energies of accelerated protons. In the case of V407 Cyg: For the weaker magnetic field  
the acceleration of protons is limited by their energy losses on pion production (see Eq.~7, thick curves), 
but  for stronger magnetic field, the acceleration is not
limited by the energy losses. (see Eq~4, thin curves). In the case of V906 Car and V339 Del, for both surface 
magnetic fields of the WD, the acceleration process is limited by their energy losses.
The scaling factor of the reconnection region is assumed to be $\eta = 1$. The neutrino spectra  are confronted 
with  the atmospheric neutrino background (ANB) within $1^\circ$ of the source collected 
during the period of 10 days (thin, dotted curves, see  Lipari~1993).
Two curves correspond the the horizontal ANB (upper flux) and vertical ANB (lower flux).}
\label{fig2}
\end{figure}
\begin{table*}
  \begin{tabular}{lllllllllll} 
\hline 
\hline 
Nova:    & d    &  $\tau$ & $L_\gamma$  & $\xi$ &  E$_{\rm Nova}$ &   $v_{\rm w}$  &  $L_{\rm w} = L_{\rm Bol}$  
& $E_{\rm max}^{\rm p}$ & $N_{\nu,det}^{\rm IceCube}$    &  Ref.      \\
         & kpc  &   days  & erg s$^{-1}$ &   $L_\gamma/L_{\rm Bol}$    &  erg  &  km s$^{-1}$ &  erg s$^{-1}$ &  TeV       &  
$\nu_\mu$ events & \\
\hline
V407 Cyg   &   2.7            &  15      &  $2.8\times 10^{35}$                &  $4\times 10^{-3}$          & $10^{44}$ &
 3200 & $7\times 10^{37}$ & 130 (207)  &  0.14  (0.16) &  (a)  \\
\hline
V906 Car  &  $4.0\pm 1.5$  &  $>20$   &  $1.7\times 10^{36}$       &  $2\times 10^{-3}$ & $1.5\times 10^{45}$ 
& 2500 & $8.5\times 10^{38}$   &  6.7 (22) & 0.125 (0.4)  & (b)\\
\hline
V339 Del   &  $4.5\pm 0.6$ &   27     &  $6.7\times 10^{34}$       &   $4\times 10^{-4}$  
& $3.9\times 10^{44}$ & 2000  &   $1.7\times 10^{38}$ & 11.5 (38) & 0.05 (0.14) & (b),(c),(d) \\
\hline 
\hline 
\end{tabular}  
\caption{Estimated muon neutrino event rates in the IceCube neutrino detector ($N_{\nu,det}^{\rm IceCube}$) from a few Novae:
d - the distance to Nova, $\tau$ - the duration of $\gamma$-ray emission phase, $L_\gamma$ - the average 
$\gamma$-ray 
luminosity, $\xi$ - the ratio of $\gamma$-ray power to bolometric luminosity $L_{\rm Bol}$, 
$E_{\rm Nova} = \tau * L_{\rm Bol}$ the total energy of Nova, $v_{\rm w}$ - the wind velocity,
$L_{\rm w} = L_{\rm Bol}$ - the wind power equal to the bolometric luminosity, $E^{\rm p}_{\rm max}$ -
the maximum energies of accelerated protons for $B_{\rm s} = 10^7$~G and $10^8$~G.   
References: (a) Abdo et al. (2010), (b) Aydi et al. (2020b), (c) Schaefer et al. (2014), (d) Shore et al. (2013b).}
\label{tab2}
\end{table*}

\subsection{A moderate Nova V407 Cyg}

V407 Cyg is the first Nova detected in the GeV $\gamma$-rays (Abdo et al.~2010). It is relatively 
nearby symbiotic Nova with a fast wind and a relatively low wind power (see Table.~1). Such features allow 
acceleration of protons in the reconnection region to  
energies above hundred TeV. We consider acceleration of protons in two cases, i.e for the surface magnetic 
field of the WD $10^7$ G and $10^8$ G. In the first case, protons 
lose efficiently energy in collisions with the matter on the pion production already in the reconnection region 
(acceleration limited by the energy losses). In the second case, acceleration of protons is not limited
by the energy losses since the maximum energies, which can be reached by protons, are below the limit given by 
Eq.~4. They are injected into a turbulent equatorial wind region. They become isotropised.
They collide with the matter producing neutrinos quasi-isotropically.  
As an example, we consider the WD with the coherence scale of the reconnection region described by $\eta = 1$.
The expected spectra of neutrinos in those two cases are shown in Fig.~2. Around TeV energies, the neutrino 
spectra from those example Nova are comparable to the ANB. 
The neutrino event rates, for the energy conversion efficiencies mentioned above and assuming isotropic 
emission, are less than one (Table~1). Therefore, their detection by the IceCube size detector
will be extremely challenging.
Note however, that in the first model (saturated acceleration), neutrino emission is 
expected to be significantly collimated. This effect can increase the neutrino event rates from specific sources 
at a cost of decreasing the chance of observability.   

\subsection{A powerful Nova V906 Car}

V906 Car is one of the most powerful Nova observed up to now in GeV $\gamma$-rays (Aydi et al 2020b). It 
belongs to 
the class of classical Novae. It has the mild wind velocity but the large wind power (see Tab.~1).
In such a case, acceleration of protons is saturated at a relatively low energies (a few to several TeVs)
for the magnetic field of the WD in the range $10^7-10^8$ G. 
The calculated neutrino spectra are below (but not very far) from the ANB (see Fig.~2). The predicted neutrino 
event rates (assuming again isotropic production) are comparable to that expected from V407 Sgr. 
Detection of such low neutrino fluxes with the IceCube is challenging.
However, in both models, 
acceleration is limited by the energy losses of protons. Therefore, neutrino emission is expected to be 
collimated within a part of the celestial sphere. 
The collimation should increase the event rates for the observer located within this solid angle.
This effect is difficult to take into account since it depends on the details of the process of proton 
acceleration and subsequent interaction.

\subsection{A weak Nova V339 Del}

V339 Del is the case of a less powerful Nova with slower wind velocity and low GeV $\gamma$-ray emission
(Ackermann et al.~2014). Protons are accelerated in 
the reconnection region to the intermediate energies (several to a few tens of TeV) for 
the range of magnetic field 
strengths $10^7-10^8$ G. For such magnetic field strengths, the acceleration process is 
limited by the energy losses. So, neutrinos should show some level of collimation which might 
increase the event rates reported in Table.~1 in the case of the observer with the neutrino emission cone. 
Predicted neutrino event rates (if re-calculated for the isotropic case) 
are a factor of a few lower than expected in the case of considered above Novae.

We conclude that the neutrino event rates (expected in this model) are below unity in the case of 
specific Novae. Their detection by the present neutrino detectors is extremely challenging unless
the neutrino flux is not strongly beamed in the direction towards the observer.
However, some muon neutrinos might be observed if the whole population of Novae (including those 
not observed in GeV $\gamma$-rays by {\it Fermi}-LAT) is 
considered. We see some effects which can increase predicted here neutrino event rates. For example, neutrino 
production is expected to be anisotropic which might increase the event rates from specific novae above unity. 
Moreover, the active phase of the proton acceleration might last longer than applied here the period of 
the GeV $\gamma$-ray 
emission. Note that, the surface of the WD shows strong soft X-ray emission. This is the evidence of a hot WD 
surface and extended activity period of the fast wind. Therefore, acceleration of protons, in the inner 
magnetosphere of the WD, may last longer than applied here activity period of the $\gamma$-ray emission.

\section{$\gamma$-rays from hadronic collisions in the wind}

$\gamma$-rays and leptons are unavoidable products of collisions of hadrons with the matter of the wind.
However, in contrast to neutrinos, $\gamma$-rays strongly interact with the soft radiation field of the Nova, 
and its dense wind, since they are produced deep within the photosphere. 
We assume that locally within the photosphere, the medium is dense enough that the radiation field can 
be locally approximated by the radiation in thermal equilibrium. Then, the local temperature of the radiation, 
as a function of distance from the WD, can be described between the WD surface and the outer radius of 
the photosphere by the condition, $L_{\rm ph} = 4\pi R^2 \sigma_{\rm SB}T^4$,  $\sigma_{\rm SB}$ is 
the Stefan-Boltzmann constant. From the recent observations 
of the Nova RS Oph, the surface temperature of the WD is estimated on $T_{\rm WD} = 7\times 10^5$ K, 
assuming the radius of WD $R_{\rm WD} = 6\times 10^8$ cm). Then, the gradient of temperature within 
the photosphere is $T(r) = 7\times 10^5/r^{1/2}$~K,
where $r$ is the distance in units of WD radius.
$\gamma$-rays, produced close to the WD surface, should be effectively absorbed in the local radiation field 
on their way outside Nova wind since the optical depth should  be very large. 
In fact, the local optical depth for $\gamma$-rays, with energies corresponding to the peak of 
the $\gamma$-$\gamma$ cross section is 
$\tau_{\gamma\gamma}\sim n_{\rm ph}R_{\rm WD}\sigma_{\gamma\gamma}\approx 870/r^{1/2}$ for $\sim$TeV 
$\gamma$-rays. 
Therefore, we conclude that those $\gamma$-rays with energies above a few GeV are completely absorbed.
On the other hand, $\gamma$-rays with energies below a few GeV can be absorbed in the matter of the wind.
The optical depth for $\gamma$-rays on this process can be estimated from
$\tau_{p\gamma} = \sigma_{p\gamma} X\sim 6L_{38}v_3^{-3}r^{-1}$, where the cross section for $\gamma$-ray 
absorption in collision with a proton is
$\sigma_{p\gamma} = 10^{-26}$ cm$^2$ (in the case of complete screening) and the column density of the matter 
in the wind is $X = R_{\rm WD}\int_{r}^\infty n(r) dr\approx 6\times 10^{26}L_{38}v_3^{-3}r^{-1}$~cm$^{-2}$. 

This absorption effects concern $\gamma$-ray photons with energies clearly above the threshold for $e^\pm$ 
pair production.
Therefore, we conclude that the GeV $\gamma$-rays, observed from Novae by the {\it Fermi}-LAT telescope, 
have to originate in another
region than considered by us reconnection of the magnetic field, e.g. either in the shock region of expanding 
Nova or in the internal shock due to collision of fast wind with slow ejecta
(see e.g. models mentioned in the Introduction). The GeV $\gamma$-ray production  has to be located outside 
photosphere of the Nova. So, in fact our calculations of the neutrino emission from Novae are not related to
their GeV $\gamma$-ray emission. Therefore, in principle also Novae not detected in  GeV $\gamma$-rays can be 
interesting targets for the neutrino telescopes.

\section{Conclusion}

We considered the production of neutrinos in collisions of relativistic protons with the matter of the 
fast wind emanating from the White Dwarf after Nova explosion. The acceleration process of protons occurs 
in the reconnection regions of the strong magnetic field of the WD driven by the dense winds. 
We show that 
protons can reach TeV energies for the parameters of the Nova explosions, as observed in the case of Novae 
showing GeV $\gamma$-ray emission. For reasonable conversion factor of energy from the Nova wind to 
the relativistic protons (of the order of $1\%$), we obtained neutrino spectra which are close to the level 
of the Atmospheric Neutrino Background. The predicted muon neutrino event rates in the IceCube detector, 
from a few specific Novae, are not so far but less than one (assuming isotropic production of neutrinos). 
Therefore, detection of predicted in this model neutrino event rates will be extremely challenging
with the current neutrino telescopes.
We argue that, for the parameters of the considered models, the acceleration of protons is
limited by their energy losses already in the reconnection region.
Then, neutrinos, from collisions of relativistic protons with the matter, should be strongly collimated 
along the direction of the reconnection region. For the location of the observer within the neutrino 
emission cone, the neutrino fluxes (and event rates in the detector from the specific Novae reported in Table.~1)
should be enhanced in respect to the isotropic case by the factor corresponding to the reciprocal of 
the solid angle in which neutrinos are emitted from the reconnection region. We conclude that Novae, 
whose reconnection regions are
accidentally directed to the observer on the Earth, might provide the  chance to produce correlated events 
in the neutrino detector of the IceCube size.    

We note, that the estimated neutrino event rate is proportional to the time scale of the activity of 
the Nova wind. 
In our example calculations, we identify this time scale with the duration of the phase of the $\gamma$-ray 
emission in Novae (typically of the order of several days). However, observations of the super-soft X-ray 
emission from the WD surface (Kahabka \& van den Heuvel~1997) indicate that the fast wind phase of the Nova 
might last longer, of the order of months. In such a case, the neutrino fluxes reported in Table.~1 will 
be proportionally increased, even by a factor of a few.

We also show that the $\gamma$-rays, produced in these same interactions as neutrinos, are not likely 
responsible for the observed GeV 
$\gamma$-ray emission from the Novae. In this scenario, the cooling process of relativistic protons occurs 
deep in the wind region of Nova, i.e. close to its hot surface and in dense wind. Therefore, $\gamma$-rays, 
from decay of pions, are expected to be completely absorbed. 
This high energy $\gamma$-ray emission requires another acceleration site of particles, e.g. in the initial 
shock wave propagating in the surrounding medium of the Nova or in 
the region of collision of the fast wind with the initially ejected material 
(see references for models mentioned in the Introduction).

\section*{Acknowledgments}
We thank the anonymous Referee for useful comments.
This work is supported by the grant through the Polish National Research Centre 
No. 2019/33/B/ST9/01904.

\section*{Data Availability}
The simulated data underlying this article will be shared on
reasonable request to the corresponding author.

%%%%%%%%%%%%%%%%%%%%%%%%%%%%%%%%%%%%

\label{lastpage}


\begin{thebibliography}{99}
\bibitem[Aartsen et al.(20170]{aa17} Aartsen, M.G. et al. 2017 ApJ 835, 151
\bibitem[Abbasi et al.(2012)]{abb12} Abbasi, R. et al. 2012 APh 35, 615
\bibitem[Abbasi et al.(2021)]{abb21} Abbasi, R. et al. 2021 in Proc. 37th International Cosmic Ray conference (Berlin, Germany)
\bibitem[Abdo et al.(2010)]{ab10} Abdo, A.A. 2010 Science 329, 817
\bibitem[Ackermann et al.(2014)]{ack14} Ackermann, M. et al. 2014, Science 345, 554
\bibitem[Ahnen et al.(2015)]{ah15} Ahnen, M.L. et al. 2015 A\& A 582, A67
\bibitem[Albert et al.~2018]{alb18} Albert, A., Andr\'e, M., Anghinolfi, M. et al. 2018 ApJ 868, L20
\bibitem[Aydi et al.(2020a)]{ay20a} Aydi, E., et al. 2020a ApJ 905, 62
\bibitem[Aydi et al.(2020b)]{ay20b} Aydi, E., et al. 2020b Nature Astronomy 4, 776
\bibitem[Babel \& Montmerle(1997)]{bm97} Babel, J., Montmerle, T. 1997 ApJ 485, L29
\bibitem[Bednarek(2021)]{bed21}  Bednarek, W. 2021 MNRAS 507, 3292
\bibitem[Chomiuk et al.(2021)]{cho21} Chomiuk, L., Metzger, B.D., Shen, K.J. 2021 ARAA 59, 391
\bibitem[Drake et al.(2021)]{dr21} Drake, J.J. et al. 2021 ApJ, submitted (ArXiv:2110.14058)
\bibitem[Fang et al.(2020)]{fa20} Fang, K. et al. 2020 ApJ 904, 4
\bibitem[Franckowiak et al.(2018)]{fra18} Franckowiak, A. et al. 2018 A\& A 609, A120
\bibitem[Friedjung(2011)]{fr11} Friedjung, M. 2011 A\&A 536, A97
\bibitem[Gagn\'e et al.(2005)]{gag05} Gagn\'e, M., Oksala, M.E., Cohen, D.H., Tonnesen, S.K., 
ud-Doula, A., Owocki, S.P., Townsend, R.H.D., MacFarlane, J.J. 2005 ApJ 628, 986
\bibitem[Gallagher \& Starrfield(1978)]{gs78} Gallagher, J.S.,  Starrfield, S. 1978 ARAA 16, 171
\bibitem[Grosse-Oetringhaus \& Reygers(2010)]{gr10} Grosse-Oetringhaus, J.F., Reygers, K. 2010 J.Phys. G 37, 3001
\bibitem[Guo et al.(2016)]{guo16} Guo, F. et al. 2016 ApJ 818, L9
\bibitem[Heck et al.(1998)]{he98} Heck, D., Knapp, J. et al., Technical Report 6019, 1998, Forchungszentrum, Karlsruhe
\bibitem[Kahabka \& van den heuvel(1997)]{kv97} Kahabka, P., van den Heuvel, E.P.J. 1997 ARAA 35, 69
\bibitem[Kato \& Hachisu(1994)]{kh94} Kato, M., Hachisu, I. 1994 ApJ 437, 802
\bibitem[Leto et al.(2006)]{le06} Leto, P., Trigilio, C., Buemi, C.S., Umana, G., Leone, F. 2006 A\&A 458, 831
\bibitem[Leto et al.(2017)]{le17} Leto, P., Trigilio, C., Oskinova, L. et al. 2017 MNRAS 457, 2820
\bibitem[Li et al.(2017)]{li17} Li, K.-L., Metzger, B. D., Chomiuk, L., et al. 2017, Nature Astronomy 1, 697
\bibitem[Lipari(1993)]{lip93} Lipari, P. 1993 APh 1, 195 
\bibitem[Martin \& Dubus(2013)]{md13} Martin, P., Dubus, G. 2013 A\& A 551,A37
\bibitem[Martin et al.(2018)]{mar18} Martin, P. et al. 2018 A\& A 612, A38 
\bibitem[Metzger et al.(2015)]{me15} Metzger, B.D. et al. 2015 MNRAS 450, 2739
\bibitem[Metzger et al.(2016)]{met16} Metzger, B. D., Caprioli, D., Vurm, I., et al. 2016 MNRAS 457, 1786
\bibitem[Nauenberg(1972)]{nau72} Nauenberg, m. 1972 ApJ 175, 417
\bibitem[Pei et al.(2021)]{pei21} Pei, S., Orio, M., Gendreau, K. et al. 2021 ATel14901 
\bibitem[Pizzuto et al.(2021)]{pi21} Pizzuto, A., Vandenbroucke, J., Santander, M. for the IceCube Collaboration 
2021, ATel 14851
\bibitem[Razzaque et al.(2010)]{raz10} Razzaque, S., Jean, P., Mena, O. 2010 PRD 82, 123012 
\bibitem[Revinius et al.(2013)]{rev13} Revinius, T., Townsend R.H., Kochukhov O. et al. 2013 MNRAS 429, 177
\bibitem[Schaefer et al.(2014)]{sc13} Schaefer, G. H., Brummelaar, T. T., Gies, D. R., et al. 2014 Nature 515, 234
\bibitem[Shore(2012)]{sh12} Shore, S.N. 2012, BASI, 40, 185
\bibitem[Shore(2013]{sh13} Shore, S. N., Skoda, P., Korcakova, D., et al. 2013, ATel5312
\bibitem[Shore \& Brown(1990)]{sb90} Shore, S.N., Brown, D.N. 1990 ApJ 365, 665
\bibitem[Sitarek \& Bednarek(2012)]{sb12} Sitarek, J., Bednarek, W. 2015  PRD 86, 063011
\bibitem[Tatischeff \& Hernanz(2007)]{th07} Tatischeff, V., Hernanz, M. 2007 ApJ 663, L101
\bibitem[Trigilio et al.(2004)]{tri04} Trigilio, C., Leto, P., Umana, G., Leone, F., Buemi, C.S. 2004 A\&A 418, 593
\bibitem[Usov \& Melrose(1992)]{um92} Usov, V.V., Melrose, D.B. 1992 ApJ 395, 575
\bibitem[Uzdensky(2007)]{uzd07} Uzdensky, D.A. 2007 (arXiv:2007.09533)
\bibitem[Vurm \& Metzger(20180]{vm18} Vurm, I., Metzger B.D. 2018 ApJ, 852, 62
\bibitem[Wagner et al.(2021)]{wag21} Wagner, S.J. et al. (HESS Collab.) 2021 ATel14844
\bibitem[Wickramasinghe \& Ferrario(2000)]{wf00} Wickramasinghe, D.T., Ferrario, L. 2000 PASP 112, 873
\end{thebibliography}
\end{document}